\documentclass[aps,prl,superscriptaddress,reprint]{revtex4-1}

\usepackage{graphicx}
\usepackage{amsmath}
\usepackage{amssymb}
\usepackage{amsfonts}
\usepackage{placeins}
\usepackage{filecontents}
\usepackage{color}
\usepackage[normalem]{ulem} 
\usepackage{soul}
\usepackage{epstopdf}
\usepackage{hyperref}
\definecolor{gold}{rgb}{0.85,.66,0}

\DeclareRobustCommand{\SkipTocEntry}[5]{}

\usepackage[sort&compress]{natbib}

\begin{document}

 \title{Nonlinear Nanophotonic Chip-space Interfaces: On-chip Generation of Structured, Topological and Spatiotemporal Lights Via Nonlinear \v{C}erenkov Radiation}

\author{Dunzhao Wei}
\thanks{These authors contributed equally.}
\thanks{weidzh@mail.sysu.edu.cn}
\affiliation{State Key Laboratory of Optoelectronic Materials and Technologies, School of Physics, Sun Yat-sen University, Guangzhou, 510275, China.}

\author{Bo Chen}
\thanks{These authors contributed equally.}
\affiliation{State Key Laboratory of Optoelectronic Materials and Technologies, School of Physics, Sun Yat-sen University, Guangzhou, 510275, China.}

\author{Shuai Wan}
\thanks{These authors contributed equally.}
\affiliation{CAS Key Laboratory of Quantum Information, University of Science and Technology of China, Hefei, 230026, China.}
\affiliation{CAS Center for Excellence in Quantum Information and Quantum Physics, University of Science and Technology of China, Hefei, 230026, China}

\author{Yixuan Wang}
\thanks{These authors contributed equally.}
\affiliation{State Key Laboratory of Optoelectronic Materials and Technologies, School of Physics, Sun Yat-sen University, Guangzhou, 510275, China.}

\author{Jiantao Ma}
\affiliation{State Key Laboratory of Optoelectronic Materials and Technologies, School of Physics, Sun Yat-sen University, Guangzhou, 510275, China.}

\author{Pi-Yu Wang}
\affiliation{CAS Key Laboratory of Quantum Information, University of Science and Technology of China, Hefei, 230026, China.}
\affiliation{CAS Center for Excellence in Quantum Information and Quantum Physics, University of Science and Technology of China, Hefei, 230026, China}

\author{Chun Chang}
\affiliation{State Key Laboratory of Optoelectronic Materials and Technologies, School of Physics, Sun Yat-sen University, Guangzhou, 510275, China.}

\author{Guixin Qiu}
\affiliation{State Key Laboratory of Optoelectronic Materials and Technologies, School of Physics, Sun Yat-sen University, Guangzhou, 510275, China.}

\author{Zelin Tan}
\affiliation{Institute for Quantum Science and Technology, College of Science, National University of Defense Technology, Changsha, 410073, China.}

\author{Xiaoshan Huang}
\affiliation{Institute of Advanced Photonics Technology, School of Information Engineering, Guangdong University of Technology, Guangzhou, 510006, China.}

\author{Yan Chen}
\affiliation{Institute for Quantum Science and Technology, College of Science, National University of Defense Technology, Changsha, 410073, China.}

\author{Tian Jiang}
\affiliation{Institute for Quantum Science and Technology, College of Science, National University of Defense Technology, Changsha, 410073, China.}

\author{Qiwen Zhan}
\affiliation{School of Optical-Electrical and Computer Engineering, University of Shanghai for Science and Technology, Shanghai, 200093, China.}
\affiliation{Zhangjiang Laboratory, Shanghai, 201204, China.}

\author{Fang Bo}
\affiliation{MOE Key Laboratory of Weak-Light Nonlinear Photonics, TEDA Applied Physics Institute and School of Physics, Nankai University, Tianjin, 300457, China.}

\author{Songnian Fu}
\affiliation{Institute of Advanced Photonics Technology, School of Information Engineering, Guangdong University of Technology, Guangzhou, 510006, China.}

\author{Xuehua Wang}
\thanks{wangxueh@mail.sysu.edu.cn}
\affiliation{State Key Laboratory of Optoelectronic Materials and Technologies, School of Physics, Sun Yat-sen University, Guangzhou, 510275, China.}
\affiliation{Quantum Science Center of Guangdong-Hong Kong-Macao Greater Bay Area, Shenzhen, 518102, China.}

\author{Chun-hua Dong}
\thanks{chunhua@ustc.edu.cn}
\affiliation{CAS Key Laboratory of Quantum Information, University of Science and Technology of China, Hefei, 230026, China.}
\affiliation{CAS Center for Excellence in Quantum Information and Quantum Physics, University of Science and Technology of China, Hefei, 230026, China}

\author{Jin Liu}
\thanks{liujin23@mail.sysu.edu.cn}
\affiliation{State Key Laboratory of Optoelectronic Materials and Technologies, School of Physics, Sun Yat-sen University, Guangzhou, 510275, China.}
\affiliation{Quantum Science Center of Guangdong-Hong Kong-Macao Greater Bay Area, Shenzhen, 518102, China.}

\date{\today}

\begin{abstract}
\noindent \textbf{Miniaturized and reconfigurable interfaces between confined optical modes within integrated photonic chips and structured light propagating in free space would serve as a cornerstone for fundamental optical science and modern photonic technology. In this work, we exploit the anisotropic nonlinear susceptibility tensors associated with thin-film lithium niobate to construct nanophotonic chip-space interfaces capable of flexibly generating and multi-dimensionally engineering structured light via injections of photons to on-chip waveguides. By harnessing the nonlinear \v{C}erenkov radiation in integrated nonlinear microring resonators, we successfully tailor the spatial profile, polarization state, emission wavelength, topological charge and temporal wave packet of structured optical vortices, exhibiting reconfigurabilities and tuning ranges far beyond the state-of-the-art. To further showcase the capabilities of our platform, we use a single pump to generate tunable optical skyrmions via the spin-orbit coupling and multi-state integrated vortex microcombs in the visible range via synergistic $\chi^{(2)}$ and $\chi^{(3)}$ nonlinear optical processes. Our work bridges the research fields of structured light and integrated nonlinear optics, providing unprecedented opportunities for spatiotemporal light generation and on-chip multidimensional nonlinear optics.}
\end{abstract}
\maketitle

\setlength{\parskip}{0.5em}
The development of nonlinear optical materials, in particular in the form of a thin film, has revolutionized integrated photonics for a broad range applications in nonlinear optics\cite{lukin2020SiC-NP,Wilson2020GaP-NP,Chang2020AlGaAs-NC,Xiang2022SiN-review-PR,Boes2023Lithium-science,Liu2023aluminum-aop,Wang2024TFLT-Nature,Chelladurai2025-TFBT-NM,Liu2025-MOS2}. The nonlinear nanophotonic devices, which feature low-power consumptions, small device footprints and unprecedented nonlinear conversion efficiencies, are now playing an increasingly important role in different branches of modern photonics, covering optical communication\cite{Xu2005Si-modulator-Nature,Wang2018LN-modulator-Nature}, frequency metrology \cite{diddams2020optical-science}, nonlinear spectroscopy \cite{Zhao2025SiNamplifier-Nature,Kuznetsov2025}, and quantum technology \cite{Nehra2022LN-squeezing-science,Li2025SiNquantum-Nature}. To date, the primary focus in building nonlinear integrated optical devices has been on controlling the degree of freedom in frequency, thanks to the innovations in dispersion engineering, phase-matching strategies, and strong mode confinements in waveguides and microresonators \cite{Dutt2024nonlinear-NRM}. Recent progress has demonstrated a previously explored paradigm, the simultaneous engineering of multiple degrees of freedom of light in an on-chip nonlinear process. By integrating nonlinear waveguides/cavities with passive photonic nanostructures, such as angular gratings and metasurfaces, $\chi^{(3)}$  platforms can now achieve guided-to-radiative mode conversion and support the emission of programmable structured optical fields \cite{Butow2024generating-NP}, spatiotemporal vortices \cite{chen2024integrated-NP,Liu2024Integrated-NP}, and entangled quantum vortices \cite{Huang2025Si-quantum-NP}. This unique combination of nonlinear optical gain and passive optical elements creates a class of on-chip devices capable of multidimensional dynamic light control across both classical and quantum regimes. However, in most existing integrated photonic devices, the nonlinear optical process and spatial field shaping remain largely decoupled: frequency conversion is typically achieved through nonlinear interactions between different optical waves, whereas spatial field modulation is implemented via linear scattering by using passive optical elements. This functionality separation inevitably brings a compromise between the efficiencies of the nonlinear process and the light scattering. One prominent example is the angular-grating-integrated microring \cite{chen2024integrated-NP,Liu2024Integrated-NP}. in which the embedded gratings enable light extraction into free space but also introduce redundant losses that reduce the efficiency of the nonlinear optical process. Instead of using passive structures to shape the spatial profile of light \cite{Butow2024generating-NP,Huang2025Si-quantum-NP}, it is feasible to tailor the structured light by exploiting the anisotropic nonlinear susceptibility tensor of the materials \cite{Tang2020BBO-NP,Zhang2021NPC-Optica}, which is still highly challenging yet elusive in an integrated platform.

In this work, we theoretically propose and experimentally demonstrate universal chip-to-free-space interfaces, in which the frequency, spatial and temporal profiles of the free-space propagating lights are tightly linked to the underlying material nonlinearities. We harness the nonlinear interactions between counter-propagating whispering-gallery modes (WGMs) confined in nonlinear microrings. This leads to out-of-plane sum-frequency generation (SFG) carrying orbital angular momentum (OAM), mediated by nonlinear \v{C}erenkov radiation (NCR) \cite{zhang2008nonlinear-PRL} . Using $x$-cut and $z$-cut thin-film lithium niobate (TFLN) microrings \cite{Honardoost2020rejuvenating}, we achieve independent and reconfigurable control over the light’s spatial profile, polarization, wavelength, OAM, and temporal wavepacket. This enables the on-demand generation of three previously inaccessible types of structured light: optical vortices with widely tunable wavelengths and topological charges (TC) \cite{Forbes2021structured-NP}, optical skyrmions with controllable skyrmion number and wavelength \cite{Shen2023skyrmions-NP}, and spatiotemporal vortex pulses with customized wavepackets \cite{Zhan2024spatiotemporal}. Our work not only introduces a previously unexplored paradigm for generating and controlling structured light with integrated photonics, but also unlocks new functionalities for state-of-the-art TFLN devices and other nonlinear thin-film platforms, offering compelling opportunities in integrated nonlinear optics \cite{Liu2022emerging-SCPMA,Dutt2024nonlinear-NRM}.

\begin{figure*}[htpb]
	\begin{center}
		\includegraphics[width=0.9\linewidth]{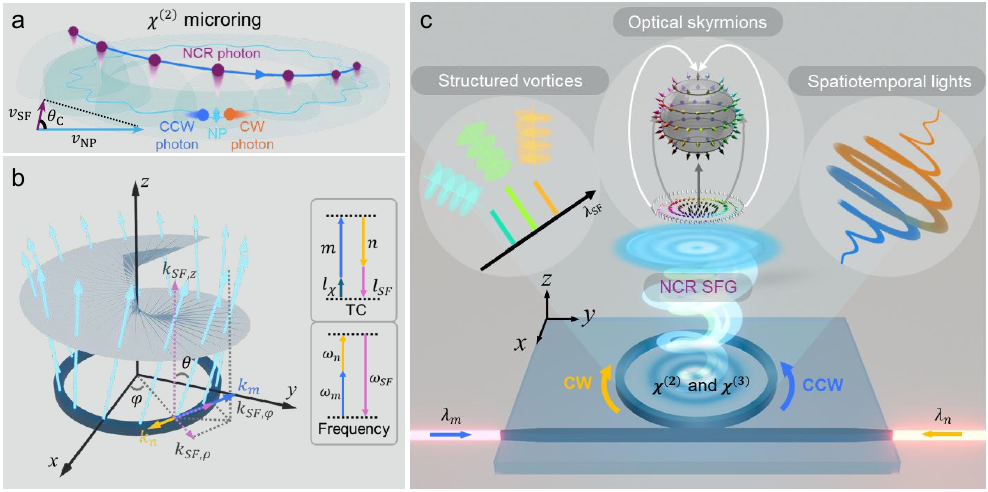}
		\caption{\textbf{Working principle of the NCR-based SF process.} (\textbf{a}) Generation of out-of-plane SF photons via the second-order nonlinear optical interaction between two counter-propagating photons in the microring governed by the NCR principle. The coherence of the SF photons provides a necessary condition to form a structured wavefront. The inset shows the NCR angle of SF field. (\textbf{b}) Theoretical analysis of the NCR SFG based on the three-wave mixing process, where wavevectors of interacting waves are introduced to analyze the structured SF wavefront. The TC and frequency of the SF field are $l_\text{SF}=m-n+l_\chi$ and $\omega_\text{SF}=\omega_m+\omega_n$, respectively.  (\textbf{c}) Generation of structured vortices with independently tunable wavelength and TC, optical skyrmions with controlled skyrmion number and tunable wavelength, and spatiotemporal vortex pulses with engineerable wave packets by tailoring the SF wave in multiple degrees of freedom. The CCW and CW WGMs are excited by the incident fields, labeled as  $\lambda_m$ and $\lambda_n$ from the left port and the right port of the coupling straight waveguide, respectively.}
		\label{fig:Fig1}
	\end{center}
\end{figure*}

\section{NCR from $\chi^{(2)}$ microring}\label{sec2}

Figure~1a shows the working principle of the integrated nonlinear chip-space interface based on NCR. In a microring  with a $\chi^{(2)}$ nonlinearity,  second-order nonlinear polarization (NP) is induced by the SF interaction of the two counter-propagating photons. The NP velocity $v_\text{NP}$ along the azimuthal direction satisfies the equation of $\frac{\omega_m}{v_m}-\frac{\omega_n}{v_n}=\frac{\omega_m+\omega_n}{v_\text{NP}}$, where $v_m$ and $v_n$ represent the phase velocities of the interacting CCW and CW photons, and $\omega_m$ and $\omega_n$ denote their frequencies \cite{boyd2008nonlinear-books}.  $v_\text{NP} =\frac{(\omega_m+\omega_n)R}{m-n}$ is obtained via harnessing the WGM resonant condition of $\frac{\omega_m}{v_m}=\frac{m}{R}$ and $\frac{\omega_n}{v_n}=\frac{n}{R}$, where $m$ and $n$ are positive integers representing the azimuthal numbers for the CCW and CW modes, respectively, and $R$ is the radius of the microring.  It suggests that $v_\text{NP}$ in the microring could significantly exceed the phase velocity of the SF photon $v_\text{SF}$, producing SF photons  with an  \v{C}erenkov radiation angle of  $\theta_\text{C}=\mathrm{arccos}(v_\text{SF}/v_\text{NP}) $, as shown in the inset of Fig. 1a .  Our configuration differs from previous NCR within bulk crystals in two aspects. First, the NP is driven by counter propagation photons, resulting in much larger $\theta_\text{C}$ to enable the SF waves couple into the free-space. Second, the NCR occurs along a curved path rather than a straight line, so that the tilted wavefront will skew into helical phase fronts, forming optical vortices \cite{cai2012integrated}. This on-chip NCR process is governed by the in-plane azimuthal-number conservation during the $\chi^{(2)}$ process.

The nonlinear wave equation should be used to quantitatively analyze the NCR SFG, which takes the following factors associated to the NP and its radiation into consideration: polarization states of WGMs, i.e., transverse electric (TE) modes or transverse magnetic (TM) modes; the $\chi^{(2)}$ tensor of the resonator material; and the diffraction induced by sub-wavelength cross-section of the microring. Figure~1b illustrates a theoretical model based on the three-wave mixing process. The electric fields of the fundamental waves are labeled as $E_{\text{P},m}$ for CCW modes and $E_{\text{P},-n}$ for CW modes, respectively. The first subscript “P” refers to a TE-like or TM-like polarization state, while the second subscript denotes the mode order, where the minus sign before “$n$” indicates the CW mode, distinguishing it from the CCW mode. The fundamental waves producing NP can be written as $E_{\text{P},m}=u_{\text{P},m}\left(\rho,z\right)e^{i\left(\omega_mt+m\varphi\right)}{\hat{e}}_\text{P}$ and $E_{\text{P},-n}=u_{\text{P},-n}\left(\rho,z\right)e^{i\left(\omega_nt-n\varphi\right)}{\hat{e}}_\text{P}$. Here, $u_{\text{P},m}\left(\rho,z\right)$ and $u_{\text{P},-n}\left(\rho,z\right)$ are the electric field distributions in the cross-section of the microring, and ${\hat{e}}_\text{P}$ is a unit vector of the dominant electric field for TE or TM modes. Therefore, the evolution of the radiated SF field can be expressed by the nonlinear wave equation \cite{berger1998nonlinear-PRL}: 
\begin{equation}\label{eq:1}
\begin{aligned}
    {\vec{k}}_\text{SF}\cdot\nabla{\vec{E}}_\text{SF}=&-i\frac{\left(\omega_m+\omega_n\right)^2}{c^2} \overleftrightarrow{\chi^{(2)}}:{\hat{e}}_\text{P}{\hat{e}}_\text{P}u_{\text{P},m}\left(\rho,z\right)\\
    &\times u_{\text{P},-n}\left(\rho,z\right)e^{i\left(m-n\right)\varphi}e^{-i\Delta \vec{k}\cdot\vec{r}}e^{i\left(\omega_m+\omega_n\right)t}
\end{aligned}
\end{equation}
where the wavevector mismatch is given by $\Delta \vec{k}=-k_{\text{SF},\rho}\hat{\rho}-k_{\text{SF},z} \hat{z}$  owing to the in-plane confinement of the fundamental waves and the satisfaction of the azimuthal-number conservation. Here, $k_{\text{SF},\rho}$ and $k_{\text{SF},z}$ are wavevector components of the SF field along the radial direction $\hat{\rho}$ and $z$-axis direction $\hat{z}$, respectively.  $\overleftrightarrow{\chi^{(2)}}:{\hat{e}}_\text{P}{\hat{e}}_\text{P}$ denotes a double contraction of the nonlinear tensor under the interaction of two unit vectors, leading to a vector effective nonlinear coefficient ${\vec{\chi}}_\text{eff}=\left[\chi_{\text{eff},x}\ \chi_{\text{eff},y}\ \chi_{\text{eff},z}\right]^{T}$. Assuming that $u_{\text{P},m}\left(\rho,z\right)$ and $u_{\text{P},-n}\left(\rho,z\right)$ are an approximately constant within the microring and the thickness $\Delta d$ of the microring is sufficiently small, the near SF field normal to the $z$-axis can be deduced from Eq. (1), given by  

\begin{equation}\label{eq:2}
\begin{aligned}
    {\vec{E}}_\text{SF-near}=&-i\frac{\left(\omega_m+\omega_n\right)^2}{k_{\text{SF},z}c^2}{\vec{\chi}}_\text{eff}u_{\text{P},m}u_{\text{P},-n}\\
    &\times e^{i\left(m-n\right)\varphi}e^{i\left(\omega_m+\omega_n\right)t}\Delta de^{i\frac{k_{\text{SF},z}{\Delta d}}2}
\end{aligned}
\end{equation}

\noindent  The far field  can be calculated by performing Fourier transformation of near field, i.e.  ${\vec{E}}_\text{SF-far} \propto \mathcal{F} \left \{ {\vec{E}}_\text{SF-near} \right \} $.

Since each component in ${\vec{\chi}}_\text{eff}$ possibly depends on the azimuthal angle $\varphi$, it can be expressed by Fourier series of $\chi_{\text{eff},j}=\sum_{l_\chi}C_{l_\chi}e^{il_\chi\varphi}$$ \left(j=x,y,z\right)$, where $l_\chi$ is an integer. Thus, the azimuthal phase term in Eq. (2) turns into $e^{i\left(m-n+l_\chi\right)\varphi}$, leading to the generation of OAM states \cite{allen1992orbital-PRA}, with a total TC of 
\begin{equation}\label{eq:3}
l_\text{SF}=l_\text{WGM}+l_\chi=m-n+l_\chi 
\end{equation}
Here, $l_\text{WGM}$ and $l_\chi$ can be seen as the modal-difference-induced TC and $\chi_\text{eff}$-induced TC, respectively. Eq. (3) is the general in-plane OAM conservation law for SFG within the nonlinear microring. Notably, the OAM conservation condition identified here has never been investigated in the context of the well known NCR in bulk nonlinear crystals, where research has traditionally focused solely on frequency conversion and conical angle control, with no consideration of spatial field properties. Replacing one of the interacting fields with an optical envelope, this theoretical model can be easily extended to describe radiated SF combs driven by an NP wave packet.  Figure~1c illustrates the schematic of our device consisting of a microring resonator and a side-coupled bus waveguide. Based on the above theoretical framework, we can engineer the structured SF wave in multi-dimensions including the spatial profile, polarization state, emission wavelength, and OAM by carefully choosing the resonant modes and effective nonlinear coefficients involved in the on-chip NCR process, while its temporal wave packet can be also engineered by a combination of $\chi^{(2)}$ and $\chi^{(3)}$ processes. 

As a proof of concept, the interactions between fundamental TE and TE modes or between fundamental TM and TM modes in $x$-cut and $z$-cut TFLN microring are studied to highlight the universality of the NCR process. Table 1 summarizes four representative processes: case 1 (or 2) for interaction between $E_{\text{TM},m}$ and $ E_{\text{TM},-n}$ (or $E_{\text{TE},m}$ and $E_{\text{TE},-n}$) modes in $x$-cut TFLN; case 3 (or 4) for interaction between $E_{\text{TE},m}$ and  $E_{\text{TE},-n}$ (or $E_{\text{TM},m}$ and  $E_{\text{TM},-n}$) modes in $z$-cut TFLN. The dominant $\chi_\text{eff}$ are rather distinct in these four cases, with some of them linked to the azimuthal angle, leading to structured sum-frequency (SF) fields with different spatial profiles and polarization states. It should be noticed that the physical coordinate system of $z$-cut TFLN is aligned with the lab coordinate system, while that of $x$-cut TFLN has a CCW rotation of $90^{\circ}$ around the $y$-axis relative to the lab coordinate system. The dominant $\chi_{31}$ and  $\chi_{33}\cos^2\varphi$ in cases 1 and 2 result in the main $x$-polarization single OAM state and superposition state, expressed as $|\text{V}\left.,m-n\right\rangle$ and $\frac{1}{2}\left|\left.\text{V},m-n\right\rangle\right.+\frac{1}{4}\left|\left.\text{V},m-n+2\right\rangle\right.+\frac{1}{4}\left|\left.\text{V},m-n-2\right\rangle\right.$ using ket notation, respectively. Moreover, case 3 produces a cylindrically oriented NP due to the dominant $\chi_{\text{eff,}x}$ and $\chi_{\text{eff,}y}$, expressed by a Jones matrix of $\left[\begin{matrix}\sin{2\varphi}&\cos{2\varphi}\\\end{matrix}\right]^{T}$, while case 4 leads to a radial NP of $\left[\begin{matrix}\cos{\varphi}&\sin{\varphi}\\\end{matrix}\right]^{T}$ due to the diffraction of the subwavelength microring. As the observed plane is restricted to the $x-y$ plane, the contribution of the $z$-component  in case 3 is significantly smaller compared to those of  $x$ and $y$-components. According to the spin-to-orbit angular momentum coupling \cite{Milione2011higher-PRL},  both cylindrical vector beams in cases 3 and 4 can be decoupled into two scalar OAM states with left-handed and right-handed circular polarizations (LCP and RCP), i.e., $\left|\left.\text{L},m-n+2\right\rangle\right.$ and $\left|\left.\text{R},m-n-2\right\rangle\right.$ for case 3 and $\left|\left.\text{L},m-n-1\right\rangle\right.$ and $\left|\left.\text{R},m-n+1\right\rangle\right.$ for case 4, respectively. We fabricate a variety of devices and customize a characterization system to demonstrate the proposed functionalities.

\begin{figure*}[htpb]
	\begin{center}
        \includegraphics[width=1.0\linewidth]{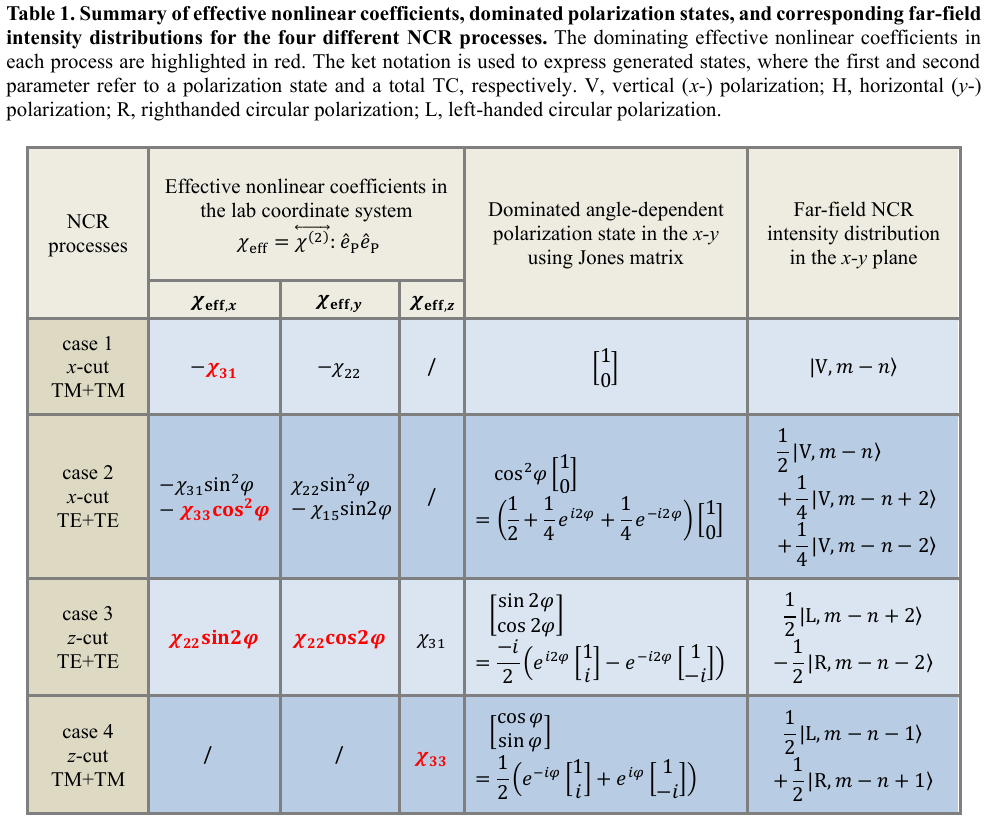}
	\end{center}
\end{figure*}
\begin{figure*}[htpb]
	\begin{center}
		\includegraphics[width=0.9\linewidth]{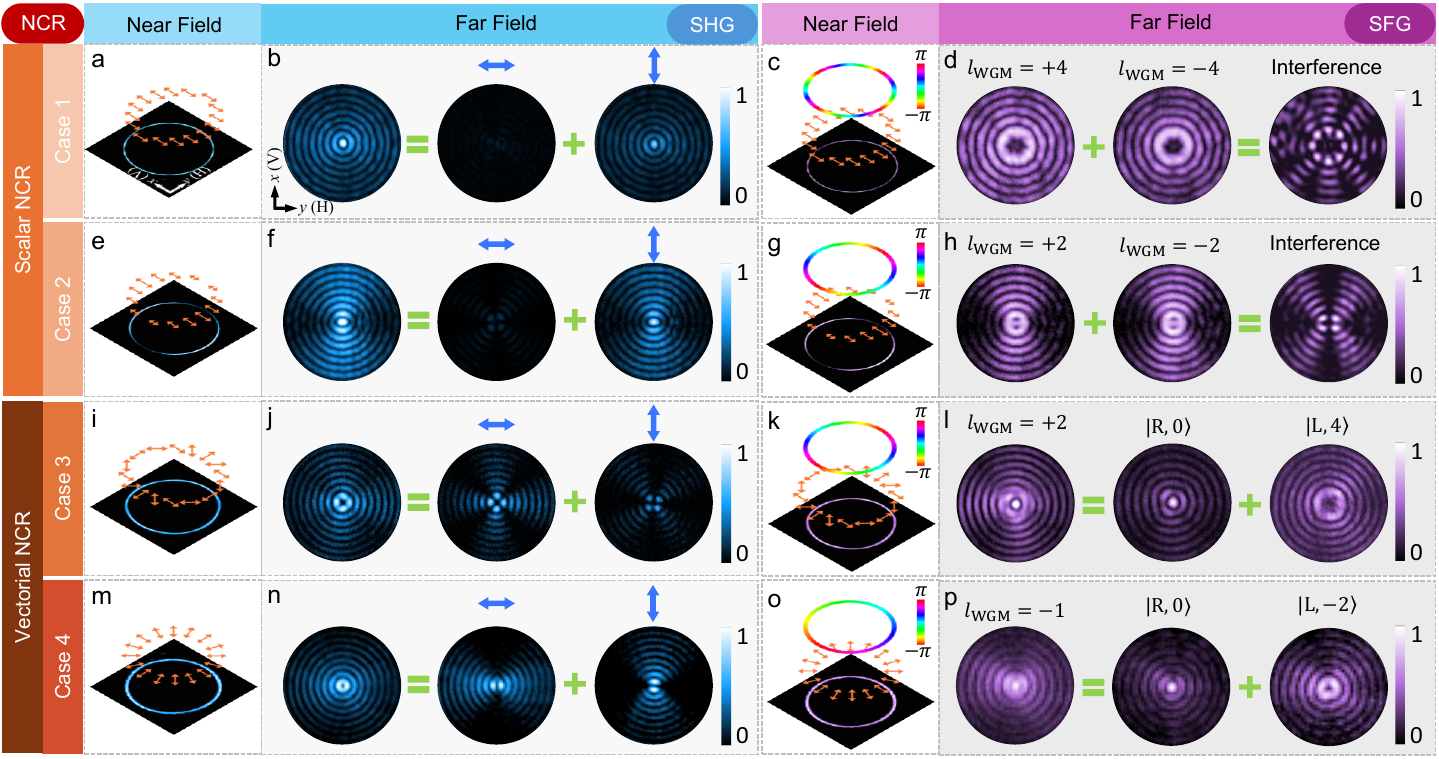}
		\caption{\textbf{Engineering on-chip structured SF fields in multiple degrees of freedom.}  (\textbf{a, e, i, m}) Measured near-field patterns of the SHG in the four cases. Homogeneous bright circular intensity patterns are observed in cases 1, 3 and 4 while amplitude modulation with two circular arcs is obtained in case 2. The arrows above the intensity patterns represent the polarization states. (\textbf{b, f, j, n}) Measured far-field patterns of SHG in the four cases with dominant $x$-polarization component in cases 1 and 2 and distinct cylindrical polarization states in cases 3 and 4.  (\textbf{c, g, k, o}) Measured near-field patterns of the SFG in the four cases, which are the same as that in SHG but carry modal-difference induced spiral phases with TCs of  $l_\text{WGM}=\ \pm4,\pm2,2, \text{and} -1$, respectively.  The arrows and colorful rings above the intensity patterns represent the polarization states and phases, respectively. (\textbf{d, h, l, p}) Measured far-field patterns of SFG with $l_\text{WGM}=\ \pm4,\pm2,2, \text{and} -1$ for cases 1 to 4, respectively, showing the scalar vortex in case 1, the scalar superposition state in case 2, and the spin-orbit coupling states in cases 3 and 4.}

		\label{fig:Fig3}
	\end{center}
\end{figure*}

\section{Generating and engineering structured SF fields in full dimensions}

In the experiment, we firstly control the spatial profile and polarization state of the SF field by setting $l_\text{WGM}=0$, and then engineer the additional wavefront by using $l_\text{WGM}\neq0$, as shown in the near-field and far-field radiation patterns in Fig.~2. $l_\text{WGM}=0$ is obtained by setting $\lambda_n=\lambda_m$, making the on-chip SFG  degenerate into the second harmonic generation (SHG). In case 1, a homogeneous second-harmonic (SH) intensity distribution on the ring is observed owing to the $\chi_{\text{eff},x}$ and $\chi_{\text{eff},y}$ being independent on the azimuthal angle (Fig.~2a). The Fourier transformation of this distribution yields a zero-order Bessel function, which aligns with the recorded far-field intensity pattern. Here, the $x$-polarization component dominates owing to the larger $\chi_{31}$, resulting in a $|\text{V}\left.,0\right\rangle$ state (Fig.~2b). Setting $\ l_\text{WGM}=m-n=\pm4$, the radiated SF field will be loaded with spiral phases and the far fields feature typical hollow beams as presented in Figs.~2c and 2d, which is in contrast to SHG without OAM. According to Table 1, these two SF vortices can be expressed as $\left|\left.\text{V},4\right\rangle\right.$ and $\left|\left.\text{V},-4\right\rangle\right.$, respectively, verified by their interference pattern with anti-node number of $2\left|l_\text{SF}\right|=8$ \cite{kulkarni2017single-shot-NC,chen2021bright}. In case 2, although the non-zero $\chi_{\text{eff},x}$ and $\chi_{\text{eff},y}$ contribute to the radiated SH field, the near-field SH intensity presents as two circular arcs due to the dominant $\chi_{33}\cos^2\varphi$ component within $\chi_{\text{eff},x}$ (Fig.~2e). This is attributed to the largest $\chi_{33}$ for TFLN. Such a circularly modulated amplitude produces OAM sidebands of $l_\chi=\pm2$ encoded by angular diffraction (Fig.~2f) \cite{Jack2008Angular-NJP}. When encoding spiral phases (Fig.~2g), these two sidebands are verified by the bright centers at the far field for states of $\frac{1}{2}\left|\left.\text{V},2\right\rangle\right.+\frac{1}{4}\left|\left.\text{V},4\right\rangle\right.+\frac{1}{4}\left|\left.\text{V},0\right\rangle\right.$ with $l_\text{WGM}=2$ and $\frac{1}{2}\left|\left.\text{V},-2\right\rangle\right.+\frac{1}{4}\left|\left.\text{V},0\right\rangle\right.+\frac{1}{4}\left|\left.\text{V},-4\right\rangle\right.$ with $l_\text{WGM}=-2$, due to the counteraction of $\ l_\text{WGM}$ and $l_\chi$ (Fig.~2h). The interference pattern for $l_\text{WGM}=\pm2$ in Fig.~2h, in contrast to the circularly symmetric pattern observed in Fig.~2d, exhibits non-uniform four nodes. In case 3, the two orthogonal polarizations in the $x-y$ plane form a cylindrical vector polarization state of $\left[\begin{matrix}\sin{2\varphi}&\cos{2\varphi}\\\end{matrix}\right]^{T}$, so that a homogeneous SH intensity in the near field is observed (Fig.~2i), but the far-field intensity pattern presents as a donut shape owing to the polarization singularity at the center (Fig.~2j) \cite{zhan2009cylindrical-AOP}. The cylindrical polarization repeating twice in one round trip is revealed by the $x$-polarization and $y$-polarization components in Fig. 2j, exhibiting a rotational angle of $45^{\circ}$ in between. Such a cylindrical vector beam can be decoupled into two circularly polarized OAM states, i.e.,$\left|\left.\text{L},l_\chi=2\right\rangle\right.$ and $\left|\left.\text{R},l_\chi=-2\right\rangle\right.$. Setting $l_\text{WGM}=$ 2 (Fig.~2k), the generated SF vortices turn into a superposition of $\left|\left.\text{L},4\right\rangle\right.$ and $\left|\left.\text{R},0\right\rangle\right.$, confirmed by far-field SF intensity patterns in Fig.~2l. The bright spot in the center for the RCP component reveals the total TC of $l_\text{WGM}+l_\chi=0$. Generally, the non-zero $\chi_{\text{eff},z}$ in case 4 cannot produce an out-of-plane SH field. However, the microring with a subwavelength width acts as a circularly symmetric diffraction structure, scattering the SH field produced by $\chi_{33}$ in the $x-y$ plane to the far field. This diffracted feature results in a radial polarization of $\left[\begin{matrix}\cos{\varphi}&\sin{\varphi}\\\end{matrix}\right]^{T}$, as shown by the homogeneous SH intensity on the ring (Fig.~2m) and the far-field cylindrical vector SH beam (Fig.~2n). It can be also expressed as a superposition state of $\left|\left.\text{L},l_\chi=-1\right\rangle\right.$ and $\left|\left.\text{R},l_\chi=1\right\rangle\right.$. Similarly, an SF state of $\frac{1}{2}\left.|\text{L},-2\right\rangle+\frac{1}{2}\left.|\text{R},0\right\rangle$ is obtained by inducing $l_\text{WGM}=-1$ (Figs.~2o and 2p). The measured spectra of the SHG and SFG processes demonstrate that the prominent out-of-plane radiation comes from nonlinear interactions of the counter-propagating two fundamental waves while the simulated far-field intensity patterns are in excellent agreement with the experimental results.

\section{Optical vortices with ultra-wide tuning ranges in both wavelength and TC}

\begin{figure*}[htpb]
	\begin{center}
		\includegraphics[width=0.9\linewidth]{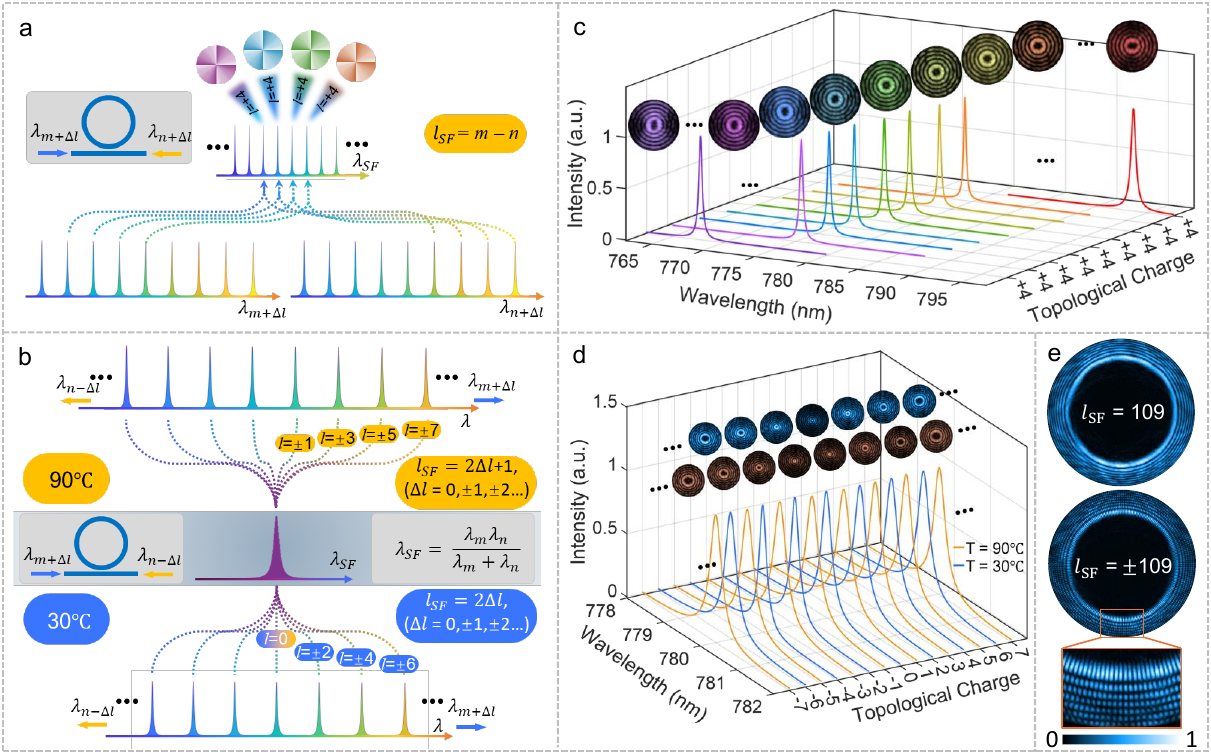}
		\caption{\textbf{Independent and wide-range tunings of the wavelength and TC of the structured optical vortices.}  (\textbf{a}) The scheme for achieving optical vortices with tunable wavelengths but a fixed TC of  $l_\text{SF}=\ m-n$ by simultaneously changing the orders of two fundamental waves along the same direction. (\textbf{b}) The scheme for achieving optical vortices with tunable TCs following the laws of $l_\text{SF}=2\Delta l$ and $l_\text{SF}=2\Delta l+1$ but at a fixed wavelength by simultaneously changing the orders of two fundamental waves along the opposite direction. Temperature control on the microring is employed to fine-tune the resonant wavelength by half of the FSR, compensating for the wavelength difference between SF vortices with odd and even TCs. (\textbf{c}) Experimental demonstrations of optical vortices with tunable wavelengths without changing the TC by keeping $m-n= 4$ in case 1 based on the scheme in (\textbf{a}). The emission wavelength spans from 767.77 nm to 793.62 nm. (\textbf{d}) Experimental demonstrations of optical vortices with tunable TCs without changing the wavelength in case 1 based on the scheme in (\textbf{b}). The TC ranges from -7 to 7 continuously and the wavelength difference between the odd and even TCs is compensated by temperature control. (\textbf{e}) The scalar SF vortex with a TC of $l_\text{SF}=$ 109 by keeping $m-n= 109$ and the on-chip interference pattern for $l_\text{SF}= \pm109$.}
		\label{fig:Fig3}
	\end{center}
\end{figure*}

We now present the ability to independently tune the SF wavelength and TC with unparalleled ranges, which significantly outperforms the state-of-the-art \cite{cai2012integrated, miao2016orbital, zhang2018inp, zhang2020tuanble-science, chen2021bright, lu2023highly, zhang2022spin-science}. In case 1 with $l_\chi=0$, optical vortices carry a single TC, rather than a superposition of TCs can be generated, as supported by the theoretical deviation presented in Table~1. Additionally, the correlation between wavelength and TC could be decoupled by simultaneously changing the orders of both fundamental waves with several free spectrum ranges (FSRs), labeled as $\Delta l$, along the same direction, i.e. $m$ to $m+\Delta l$ and $n$ to $n+\Delta l$, or opposite direction, i.e. $m+\Delta l$ and $n-\Delta l$. The former could produce optical vortices with a pre-designed TC, i.e., $\left.|\text{V},m-n\right\rangle$, but varied emission wavelengths (Fig.~3a), while the latter can tune the TC of OAM beams, i.e., $\left.|\text{V},m-n+2\Delta l\right\rangle$, with a targeted emission wavelength (Fig.~3b). We set $\lambda_m=$ 1562.02 nm and $ \lambda_n=$ 1556.55 nm to produce OAM with $l_\text{SF}=4$ at first, and then tune the pump wavelengths to shorter and longer values, achieving the lowest $\Delta l=-17$  and highest $\Delta l= 20$, respectively. This tuning process covers a tuning bandwidth of about 51.8 nm, corresponding to 37 times of the FSR (1.4 nm for each FSR). As a result, the nearly unchanged far-field SF intensity patterns indicate the invariable TC while the emission wavelength spans from 767.77 nm to 793.62 nm, as shown in Fig.~3c. We note that the asymmetries of the far-field patterns at the shortest and longest wavelengths can be attributed to interleaving and even couplings between TE and TM mode families with unequal FSRs in the $x$-cut TFLN microring \cite{Pan2019}. To demonstrate the independently tunable TC, the fundamental wavelengths are first fixed at $\lambda_m=\lambda_n=$ 1562.02 nm, and then they are tuned to obtain an even TC of $l_\text{SF}=2\Delta l$. In this configuration, the TC changes from -6 to 6 with a step of 2 at the same SF wavelength of 781.01 nm, as the blue spectra and patterns shown in Fig.~3d.   Setting $\lambda_m=$ 1560.65 nm and $\lambda_n=$ 1562.03 nm to get $m-n=1$ with an SF wavelength of 780.66 nm, the odd TC follows the law of $l_\text{SF}=2\Delta l+1$. It should be noticed that there is a slight difference between the SF wavelengths for odd and even TCs. We used temperature control to tune the resonant wavelength of the microring by half an FSR when producing optical vortices with odd TCs, so that the SF wavelength stays the same as that with even TCs. The orange spectra and far-field patterns in Fig.~3d show the OAM with odd TCs ranging from -7 to 7 after wavelength tuning. In the extreme case, when setting $\lambda_m=$ 1481.35 nm and $\lambda_n=$ 1629.42 nm, the optical vortex achieves a TC as high as $l_\text{SF}=109$, whose TC value is verified through the multiple anti-node pattern obtained by its interference with the OAM state carrying the TC of $l_\text{SF}=-109$, as shown in Fig.~3e. The demonstration of independently tuning the emission wavelength of 37~FSR (51.8~nm) and the TC over 100, orders of magnitude wider than the existing devices, provides a powerful tool for high-dimensional optical information processing \cite{Cozzolino2019,Willner2021}.

\begin{figure*}[htpb]
	\begin{center}
		\includegraphics[width=0.9\linewidth]{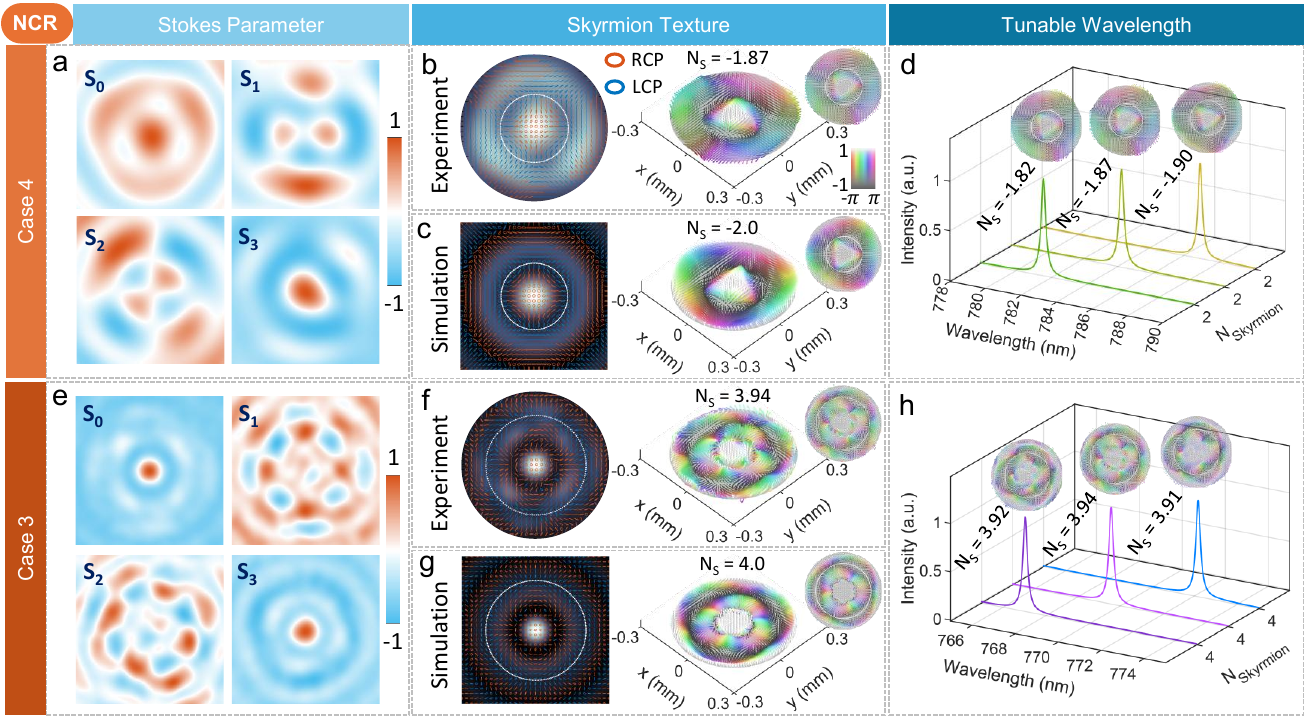}
		\caption{\textbf{Wavelength-tunable optical skyrmions with skyrmion numbers of -2 and 4.}  (\textbf{a}, \textbf{e}) Measured Stokes parameters in case 4 with $m-n=1$ and case 3 with $m-n=2$, respectively. (\textbf{b}, \textbf{f}) Recovered polarization distributions based on (\textbf{a}, \textbf{e}) and corresponding spin textures mapped from a unit Poincaré sphere onto a confined plane. Skyrmion numbers of -1.87 and 3.94 are experimentally extracted, respectively. (\textbf{c}, \textbf{g}) The theoretical polarization distributions and vortex textures for optical skyrmions with skyrmion numbers of -2 and 4, respectively. (\textbf{d}, \textbf{h}) The tunable wavelength of the optical skyrmions in (\textbf{b}, \textbf{f}) by simultaneously changing the orders of both fundamental waves along the same direction with $\Delta l= \pm2 $.}
		\label{fig:Fig4}
	\end{center}
\end{figure*}

\section{Optical skyrmions with controllable skyrmion number and wavelength}
Optical skyrmions, first realized through nanophotonic spin-orbit interaction in evanescent waves \cite{Tsesses2018,Du2019}, hold transformative potential for applications in data storage, communication, imaging, and quantum technologies due to their unique topologically stable textures \cite{Shen2023skyrmions-NP}. In particular, on-chip optical skyrmionic emitters are highly desirable for technological advancements yet their realization is highly elusive. The first on-chip skyrmionic beam generator was very recently demonstrated in a silicon micoring resonator with two sets of angular gratings \cite{Lin2024}. Alternatively, we can directly generate optical skyrmions by employing NCR-based SFG and avoid the implementation of angular gratings, offering reconfigurability in skyrmion texture and tunability in emission wavelength. The creation of optical skyrmions involves the superposition of two OAM modes with distinct TCs and opposite circular polarizations. This superposition creates a spatial distribution of pseudospin vectors that can be mapped onto a unit Poincaré sphere, effectively unwrapping the skyrmion structure \cite{Shen2022generation}. Consequently, the superposition states illustrated in Figs.~2l and 2p are capable of producing skyrmions with skyrmion numbers of $\pm 2$ and $\pm 4$, respectively. In the experiment, we first characterize the optical skyrmion in case 4 with $m-n = 1$. By measuring the four Stokes parameters of the radiated SF field (Fig.~4a), we unveil the polarization distribution across the total intensity patterns (the left panel of Fig.~4b). The reconstructed polarizations can be fully mapped by the pseudospin vectors in the planes transformed from a unit Poincaré sphere, as shown in the right panel of Fig.~4b, from which a skyrmion number of -1.87 is extracted. For comparison, the simulated polarization distribution and the spin texture based on Table 1 are also presented in Fig.~4c. Figure~4d demonstrates the flexible wavelength tuning of this optical skyrmion by simultaneously changing the orders of both fundamental waves along the same direction with $\Delta l= \pm2 $, with the measured skyrmion number remaining stable and close to the theoretical value of -2. Similarly, we analyze the Stokes parameters in case 3 with $m-n = 2$ (Fig.~4e), obtaining the polarization distribution and vortex texture (Fig.~4f). The measured skyrmion number is 3.94, close to the theoretical value of 4 (Fig.~4g). It also features tunable wavelength and a good skyrmion number, as shown in Fig.~4h. Skyrmion numbers of 2 and -4 can be obtained by setting $m-n = -1$ in case 4 and $m-n = -2$ in case 3, respectively. Full skyrmions with skyrmion numbers of -2 in Fig. 4c and 4 in Fig. 4g are obtained by confining the SF signals within the first and second nodes of the zeroth-order Bessel beam, respectively, where the pseudospin vectors cover the entire Poincaré sphere and optical fields contain most of the power \cite{Shen2022generation}. These findings bring on-chip optical skyrmions into the nonlinear optics regime and endow the power of reconfiguring the skyrmion texture and tuning wavelength independently, beyond the device performances provided by linear optics. Moving forward, it would be very exciting to explore, with our platform, quantum entanglement in nonlinear optical skyrmions, which have been recently shown with spontaneous parametric down-conversion  sources and plasmonic structures\cite{quantumskyrmions}.

\begin{figure*}[p]
	\begin{center}
	\includegraphics[width=0.7\linewidth]{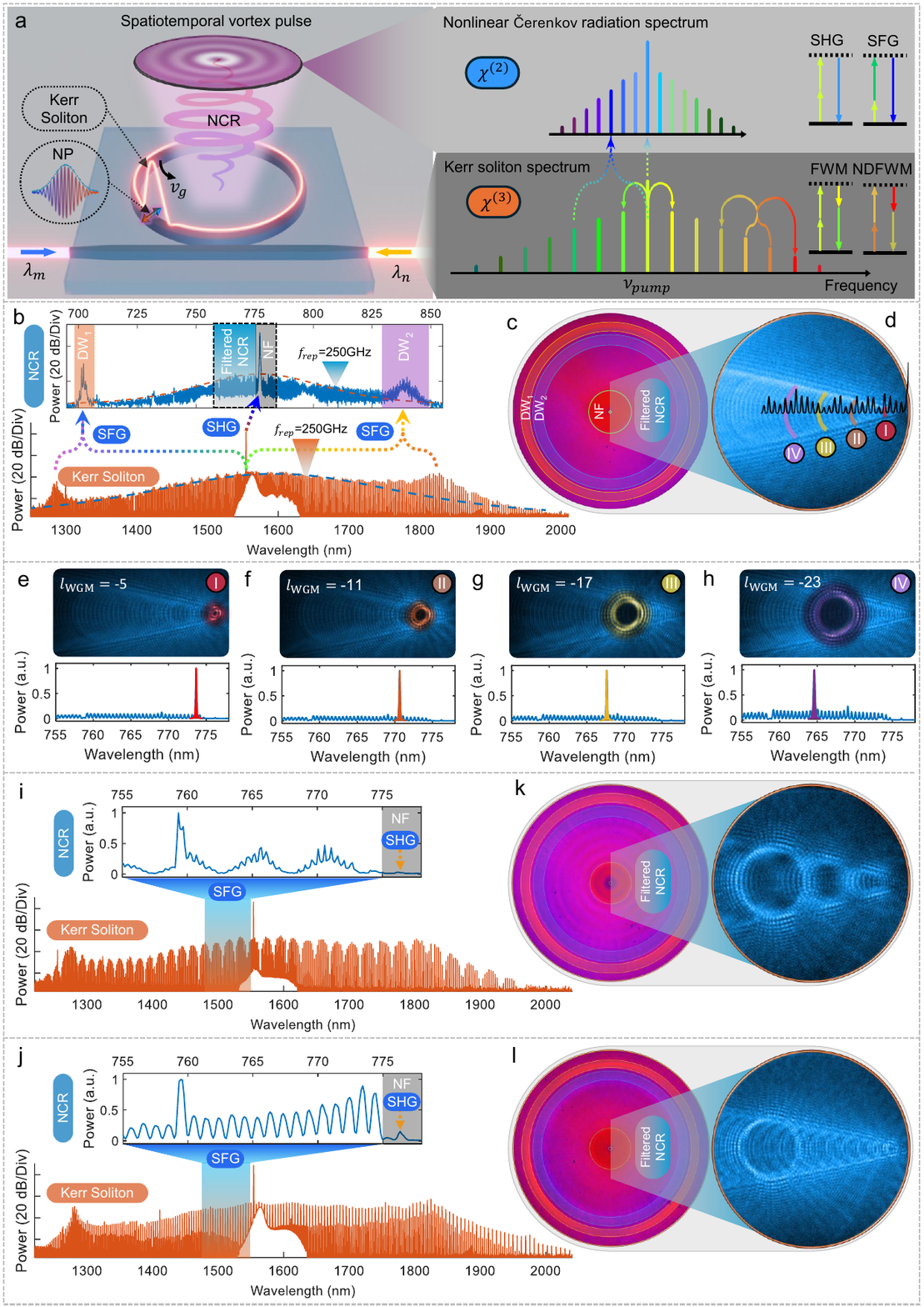}
	\caption{\textbf{Spatiotemporal vortex pulses with engineerable wave packets.} (\textbf{a}) A schematic illustrating the principles of  $\chi^{(3)}$ and $\chi^{(2)}$ nonlinear interactions that produce telecom-band Kerr solitons and short NIR spatiotemporal SF vortex. The telecom-band Kerr solitons are generated by the CCW pump via four-wave mixing processes while the short NIR spatiotemporal SF vortex is produced from the circulating NP packet formed by the interactions between the CCW soliton and the CW continuous pump at the group velocity of $v_{\text{g}}$ but with a superluminal phase velocity along the azimuthal direction. (\textbf{b}) Spectrum of a Kerr soliton with the repetition rate of 250 GHz (2 nm in wavelength) in the telecom band and spectrum of the corresponding short NIR spatiotemporal SF vortex with the same repetition rate of 250 GHz (0.5 nm in wavelength) in short NIR. A notch filter (gray area denoted as NF) is used to suppress the SH spectrum from the CCW and CW pumps and a bandpass filter (blue area denoted as filtered spectral lines) is employed to spectrally select the spectral lines of interest (with low TCs). The dispersive waves at short (orange area) and long (purple area) wavelengths are marked by DW$_1$ and DW$_2$, respectively. (\textbf{c}) Radiated intensity profiles of spatiotemporal SF vortex. The far-field patterns associated with the vortices with different OAMs selected by the bandpass filter and dispersive waves correspond to the green, blue and orange rings, respectively. (\textbf{d}) The filtered spatiotemporal SF vortex is dispersed by a blazed grating, resulting in different diffraction angles for the spectral lines at different wavelengths. The zoomed-in comb lines at different wavelengths in the spectrum overlay the corresponding diffractive rings (I, II, III, IV) at different angles. (\textbf{e-h}) Amplified single NIR spectral lines of spatiotemporal SF vortex and corresponding vortex far-field patterns (I, II, III, IV) by tuning an extra CCW pump mimicking the targeted soliton comb line to interact with the CW pump.   (\textbf{i}, \textbf{j}) The spectra of the dual-soliton state and soliton-crystal state generated in the telecom band. Above them are the zoomed-in spectra of the up-converted spatiotemporal SF vortex in the short NIR range. (\textbf{k}, \textbf{l}) are the diffracted far-field patterns associated with the spatiotemporal SF vortex with different temporal wave packets. The clear intensity modulations respective to (\textbf{d}) reveal the characteristics of a dual-soliton state with a modulated envelope and a soliton-crystal state with doubled mode spacing.}
		\label{fig:Fig5}
	\end{center}
\end{figure*}

\section{Spatiotemporal vortex pulses with engineerable wave packets}

Finally, we go beyond continuous SF wave by using ultra-short optical pulses to create a broadband spatiotemporal SF vortex. Instead of employing external pulses from a commercial oscillator, we directly generate CCW Kerr solitons in the microring with a continuous laser, thanks to the $\chi^{\left(3\right)}$ nonlinearity\cite{wang2019monolithic-NC,He2019self-starting-optica,Gong2020Near-octave-optica,wan2024photorefraction-LPR}. In this case, the CCW optical pulse (Kerr soliton) interacts with the CW continuous wave pump forming a NP wave packet that circulates together with the soliton at the same group velocity and produces a SF spatiotemporal vortex pulse, as schematically illustrated in Fig.~5a. 
We collect the SF spectra in free space and the soliton pulses in the waveguide, as presented in Fig.~5b. The smooth envelope together with the two dispersive waves of soliton propagating in the microring are imprinted in the spatiotemporal SF field traveling in the free space. To reveal OAM information encoded in the individual spectral line, a notch filter (NF) is used to suppress the up-converted pump (grey area) while a bandpass (BP) filter is used to spectrally select the spectral lines with small TCs (blue area) as marked in Fig.~5b. The far-field intensity distribution of the spatiotemporal SF vortex is presented in Fig.~5c, in which the dispersive waves and filtered comb lines are marked by circles with varied diameters due to their significant differences in the OAM states. The filtered spatiotemporal vortex, denoted by the filtered comb region in Fig.~5c is further dispersed by a blazed grating in order to spatially separate the different spectral lines, as shown in Fig.~5d, overlaying with the spectrum of the filtered SF field. The diameter of each ring increases from right to left, indicating a larger $l_\text{WGM}$ for the shorter wavelength spectral line, as exemplified by I to IV. Such a wavelength and TC correlation is further verified by introducing an additional weak fundamental beam that mimics the single spectral line of the Kerr soliton. Its wavelength can be finely tuned to align with any of the Kerr comb lines (e.g., I to IV), thereby amplifying the corresponding SF spectra and the OAM states. As shown in Figs.~5e-h, the additional fundamental CCW beam is set to match higher resonant modes with 5 (I), 11 (II), 17 (III), and 23 (IV) FSRs from the $\lambda_n=1554.45$\ nm, resulting in the enhanced SF signals with $l_\text{WGM}=-5, -11, -17, \mathrm{and} -23$ at different SF wavelengths in both spectra and the corresponding far-field patterns. In addition, we also generate a dual-soliton state \cite{cole2017soliton-NP} featuring a modulated envelope (Fig.~5i) and a soliton crystal state (Fig.~5j)  \cite{karpov2019dynamics-NPhy} exhibiting a doubled FSR at telecom band via a $\chi^{\left(3\right)}$ process, and therefore produce SF vortices with different temporal packets, as shown on the top of Figs.~5i and 5j. As clearly demonstrated in Figs.~5k and 5l, the modulated envelope in the dual-soliton state and the doubled mode spacing in the soliton crystal state are revealed in their far-field SF intensity distributions and the mode spaces. Our scheme provides an alternative to overcome the challenges of strong material dispersion and high Raman gain for on-chip generation of short NIR soliton microcombs \cite{Lu2023}. Compared to pioneering studies on spatiotemporal vortices in the telecom spectrum \cite{Liu2024Integrated-NP,chen2024integrated-NP}, our work demonstrates, for the first time, on-chip spatiotemporal vortices operating in the near-infrared spectrum.

\section{Conclusion} 
We have successfully demonstrated chip-space interfaces on an integrated nonlinear optical chip, which is capable of converting confined optical modes in waveguides to free-space propagating structured lights. By exploiting the anisotropic nonlinear susceptibility tensors to engineer SF field in multiple degrees of freedom, three types of novel states of light, that were previously inaccessible, are successfully generated, covering structured optical vortices with ultra-wide tuning ranges of wavelengths and OAM, optical skyrmions with controllable skyrmion number and tunable wavelength and spatiotemporal vortex pulses with engineerable temporal wave packets. We emphasize that the advantages associated with the NCR SFG presented in our work can be directly transplanted to other popular nonlinear optical platforms with $\chi^{\left(2\right)}$ nonlinearities, such as AlN, SiC, GaP and AlGaAs for integrated nonlinear photonics \cite{pu2016AlGaAs-optica,Wilson2020GaP-NP,Chang2020AlGaAs-NC,lukin2020SiC-NP, Liu2023aluminum-aop}. From the viewpoint of fundamental physics, the virtual non-relativistic circulating nonlinear polarizations in the microring mimic the behaviors of the accelerating charged particles in synchrotron facilities, potentially serving as an analogue simulator to shed light on high-energy physics \cite{Henstridge2018-SR-Science}. For practical applications, the on-chip NCR can be used as a diagnostic tool for analyzing the nonlinear properties of thin films, including characterizing the periodically poled structures  in TFLN \cite{Lu2020-PPLN}, determining the structure symmetry of $\chi^{\left(2\right)}$ thin films \cite{Kuo2014}, and imaging the self-poling $\chi^{\left(2\right)}$  gratings in SiN \cite{Lu2020-sin,Nitiss2022}. Our work bridges SFG to structured light, topological optics, and integrated photonics, bringing an entirely unexplored paradigm for chip-scale multidimensional nonlinear optics. Since  transverse phase matching based on $\chi^{\left(3\right)}$ nonlinearity has been demonstrated and employed in the development of  nonlinear near-filed optical microscopy \cite{Frischwasser2021}, we expect that a  $\chi^{\left(3\right)}$-based NCR effect may be uncovered in future work.

\bibliographystyle{naturemag}
\bibliography{NatureBib-new}

\vspace{0.8em}\noindent \textbf{Acknowledgements}
\noindent {This research is supported by the National Key R\&D Program of China (2021YFA1400800), the National Natural Science Foundation of China (62035017, 12361141824, 12274474, 12293052, 92250302, 12434012, 92050202), the Natural Science Foundation of Guangdong (2022B1515020067, 2023B1515120070, 2024B1515040013), Guangdong Introducing Innovative and Entrepreneurial Teams of “The Pearl River Talent Recruitment Program” (2021ZT09X044), Innovation Program for Quantum Science and Technology (2023ZD0300804). We thank Changlin Zou for the insightful discussion, Juntao Li and Haowen Liang for loaning the equipment. This work was partially carried out at the USTC Center for Micro and Nanoscale Research and Fabrication.}

\vspace{0.8em}\noindent \textbf{Author Contributions}

\noindent J.L. and D.Z.W.conceived the project. D.Z.W., C.C., X.H.W. and J.L. developed the theory. B.C., D.Z.W., S.W. and Y.X.W. performed the numerical simulations. S.W., P.Y.W., Z.L.T. and Y.C. fabricated the devices. D.Z.W., B.C., S.W., Y.X.W., J.T.M. and J.L. built the setup and characterized the devices. D.Z.W., B.C., S.W., Y.X.W., J.T.M., G.X.Q., X.S.H, T.J., S.N.F., C.H.D. and J.L. analyzed the data. D.Z.W. and J.L. wrote the manuscript with inputs from all authors. D.Z.W., S.N.F., Q.W.Z., F.B., X.H.W., C.H.D. and J.L. supervised the project.

\vspace{0.8em}\noindent \textbf{Competing interests}
{The authors declare no competing financial interests.}

\vspace{0.8em}\noindent \textbf{Data and materials availability}
{The data sets will be available upon reasonable request.}

\end{document}